\documentclass{article}
\addtocounter{equation}{0}
\begin{document}
\title{Gravitational field of a spinning cosmic string}         % Enter your title between curly braces
\author{E. \v{S}im\'{a}nek \footnote {Electronic address: simanek@ucr.edu}\\Department of Physics, University of California, Riverside, CA 92521}      % Enter your name between curly braces
\date{}          % Enter your date or \today between curly braces
\maketitle

\begin{abstract}

We study the effect of internal space rotation on the gravitational properties of infinite straight and stationary cosmic strings.  From the approximate solution of Einstein equations for the spinning Q-lump string, we obtain long range gravitational acceleration resembling that of a rotating massive cylindrical shell.  We also compute the angular velocity of the inertial frame dragging and the angle of light deflection by the Q-lump string. Matter accretion onto spinning strings can play a role in galaxy formation when the angular velocity times the string width is comparable to the speed of light.

\end{abstract}

PACS number(s):  11.27.+d, 12.39.Dc, 98.80.Cq, 98.62.Ai
\\[10pt]

\section{Introduction}

	Cosmic strings are one of a family of topological defects generated by phase transitions in the early universe $^{1,2}$.  The gravitational field of strings is of particular interest in view of their possible role in the galaxy formation$^{2}$. Vilenkin$^{3}$ studied an infinite straight static string and found that its gravitational field is very different from that of a massive rod.  First of all, there is no net active gravitational mass associated with this string.  This is because the tension in the string leads to a negative active gravitational mass which balances the positive mass of the energy density.  Thus a test particle in the neighborhood would experience no gravitational force.  Consequently, straight static strings would not initiate the gravitational instability necesary for galaxy formation. Nevertheless, owing to the unusual geometry of their exterior metric, they are of considerable astronomical importance.  Vilenkin$^{3}$ finds that the metric exhibits a geometry of a conical space with a wedge removed, and concludes that the string acts as a gravitational lens.

The absence of the active gravitational mass for a static string, parallel to the x$^{3}$-axis,  is linked to independence of string fields of $x^{0}$ and $x^{3}$.  This leads to the invariance of the stress-energy tensor $T_{\mu}^{\nu}$ with respect to the Lorentz boost in the x$^{3}$-direction.  It follows that $T^{0}_{0} = T^{3}_{3}$, an equality which makes the active mass density vanish.  However, if the string solution is time dependent, the Lorentz boost invariance is lost and a finite mass density ensues.  For the Abelian Higgs model, the string solutions are fully characterized by their position and there is no rotational degrees of freedom.  Hence, unless the string moves, the Abelian Higgs field remains time independent.  On the other hand, a non Abelian model provides a degree of freedom corresponding to rotations in the internal space.  One particularly elegant non Abelian model allowing a classical internal spin is the sigma model-lump due to Leese$^{4}$.  The solutions of this model were termed Q-lumps where $Q$ is the charge of the model. Owing to the spinning in the internal phase, the Q-lump solutions are time dependent but their position as well as energy remain static.  Thus, a stationary cosmic string built from the Q-lump solution may provide a source of an active gravitational mass.  This contrasts with the absence of the active mass for static sigma model string.$^{5}$

The purpose of the present work is to investigate the gravitational properties of the Q-lump string in the weak field approximation. In Secs. 2 and 3, we derive the long range gravitational acceleration and show that, for distances well beyond the string width, the string acts as a massive rod.  Sec. 4 is devoted to the dragging of the inertial frame caused by the spinning.  This effect has been investigated for rotating spherical mass shells since 1918 and it carries a name "Lense-Thirring" effect$^{6-8}$.  In Sec. 5, we study the deflection of light ray propagating in the plane perpendicular to the string axis.  Finally, in Sec. 6 we discuss some cosmological consequences of the present work.  

\section{Lagrangian and Stress-Energy Tensor}

We consider an infinite, straight and stationary string parallel to the x$^{3}$-axis.  The Lagrangian density is obtained by trivially extending the 2+1 theory$^{4}$ of the Q-lump to 3+1 dimensions.  Since the fields of our model are independent of x$^{3}$, the added dimension does not change the original form of the Lagrangian density of the Q-lump$^{4}$.  The basic field is a triplet $\vec{\phi}\equiv(\phi_{1}, \phi_{2},\phi_{3})$ of real scalar fields satisfying the constraint $\vec{\phi} . \vec {\phi} = 1 $.  We write the Lagrangian density in the form $(\hbar = c =1)$ 

\begin{equation}\label{Eq1}
\pounds = \eta^{2} \Bigr[\frac {1}{4} g ^{\mu \nu} \partial_{\mu} \vec{\phi}. \partial_{\nu} \vec{\phi} - \frac{\alpha^{2}}{4} (1 - \phi^{2}_{3})\Bigr]
\end{equation}

The constant $\eta^{2}$ is proportional to the energy per unit length of the pure sigma model string$^{5}$ whose Lagragian follows from (1) by putting $\alpha = 0$.

The second term in Eq. (1) is reminiscent of the uniaxial anisotropy energy of a nonrelativistic Heisenberg ferromagnet where it is responsible for a Larmor precession of the magnetization vector $\vec{M}$ about the x$^{3}$-axis.  In that case, the equations of motion for $\vec{M}$ have localized solutions in 2+1 dimensions called precessional solitons$^{9}$.  In the absence of an external magnetic field, the precessional frequency of these solitons is given by the ferromagnetic resonance frequency proportional to the anisotropy field$^{9}$ (given by $\alpha^{2}$ in the present notation).  The precessional soliton resulting from Eq. (1) is, however, of different nature.  Solving the equations of motion, Leese$^{4}$ finds precessional frequency equal to $\alpha$.  This stems from the fact that the dynamics of the Lagrangian (1) is second order in the time derivatives of $\vec{\phi}$ whereas it is first order for the Lagrangian for the nonrelativistic ferromagnet$^{9}$.  Note that the sign of the potential energy in Eq. (1) is opposite to that seen in Eq. (2.1) of Ref.[4].  This is due to our choice of the space-time metric which in the flat space limit has signature (1,-1,-1,-1)(see also Eq. (6.70) in Manton and Sutcliffe$^{10}$).

Anticipating nonzero gravitational density for the spinning string, we write the metric in the form

\begin{eqnarray}\label{Eq2}
ds^{2}= g_{00}(r)(dx^{0})^{2}- \Omega^{2}(r)\Bigr[(dx^{1})^{2}+ (dx^{2})^{2}\Bigr] \nonumber \\* +2 \Bigr[g_{01}(\vec{x}) dx^{0} dx^{1} + g_{02}(\vec{x})dx^{0} dx^{2}\Bigr] + g_{33}(r)(dx^{3})^{2}
\end{eqnarray}

where $\vec{x}$ is a vector in the $x^{1}, x ^{2}$ space, and $r = [(x^{1})^{2} + (x^{2})^{2}]^{\frac{1}{2}}$. 

The metric for the static string of the pure $(\alpha = 0)$ sigma model$^{5}$, is recovered from Eq. (2) by letting $g_{00} = 1$, $g_{01}= g_{02} = 0$, and $g_{33}= - 1$.

The components of our metric tensor $(g_{00}, g_{11}= g_{22} = - \Omega^{2}, g_{33}, g_{01}$ and $ g_{02})$ are obtained by solving the corresponding components of the Einstein equation$^{11}$.

\begin{equation}\label{Eq3}
R_{\mu \nu} = - 8 \pi G (T_{\mu\nu} - \frac{1}{2} g_{\mu\nu} T) 
\end{equation}

where $R _{\mu\nu}$ is the Ricci tensor, $T_{\mu\nu}$ is the stress-energy tensor and $T = T ^{\mu}_{\mu}$.  Using Eq. (1), we have 

\begin{equation}\label{Eq4}
T_{\mu\nu} = \frac {2}{\sqrt {| g |}} \frac {\partial}{\partial g^{\mu\nu}} (\sqrt {|g|} \pounds) = \frac{1}{2} \eta^{2} \partial_{\mu}\vec{\phi}. \partial_{\nu} \vec{\phi} - g_{\mu\nu} \pounds
\end{equation}

where $g$ is determinant of the metric.  In what follows, we treat the equation (3) in the weak field limit which amounts to replacing $g_{\mu\nu}$ in Eq. (4) by $g_{\mu\nu}=$ diag$ (1,-1,-1,-1)$.  It is convenient to calculate $T_{\mu\nu}$ by discarding the field $\vec{\phi}$ in favor of a complex scalar field $u (t,z)$ where $z = x^{1}+ i x^{2}$, and

\begin{equation}\label{Eq5}
u = \frac{\phi_{1} + i \phi_{2}}{1 + \phi_{3}} 
\end{equation}

This leads to the CP$^{1}$ representation in which the Lagrangian density of Eq. (1) becomes $^{4,10}$

\begin{equation}\label{Eq6}
\pounds = \eta^{2} \Bigr [\frac{1}{2} g ^{\mu\nu}\frac{\partial _{\mu}u \partial _{\nu} \overline{u} + \partial_{\mu}\overline{u} \partial_{\nu}u}{(1 + |u|^{2})^{2}} - \frac{\alpha^{2} u \overline{u}}{(1+ |u|^{2})^{2}}\Bigl]
\end{equation}

For the stress tensor (4), we obtain in this representation

\begin{equation}\label{Eq7}
T_{\mu\nu} = \eta^{2} \frac{\partial_{\mu}u \partial_{\nu} \overline{u} + \partial_{\mu}\overline{u}\partial_{\nu}u}{(1 + |u|^{2})^{2}} - g_{\mu\nu} \pounds
\end{equation}

\section{Active Gravitational Mass}

To calculate the active gravitational mass, we take the 00-component of Eq. (3).  We write $g_{\mu\nu}= \eta_{\mu\nu} + h_{\mu\nu}$ where $\eta_{\mu\nu}=$ diag $(1,-1,-1,-1)$.  In the weak field limit, we have 

\begin{equation}\label{Eq8}
R_{00}= - \frac{1}{2}(\partial ^{2}_{1} + \partial ^{2}_{2}) h_{00} = - \frac{1}{2} \bigtriangledown^{2}h_{00}
\end{equation}

On the right hand side (rhs) of Eq. (3) we need the expression $T_{00} - \frac{1}{2} T$.  Using Eqs. (6) and (7), we obtain

\begin{equation}\label{Eq9}
T_{00} = \frac{\eta^{2}}{(1 + |u|^{2})^{2}} (\partial_{t}u \partial_{t}\overline{u} + \partial_{i} u \partial _{i} \overline{u} + \alpha^{2}u \overline{u})
\end{equation}

and 

\begin{equation}\label{Eq10}
T_{33}= \frac{\eta^{2}}{(1 + |u|^{2})^{2}}(\partial_{t}u \partial_{t}\overline{u} - \partial_{i} u \partial_{i}\overline{u} - \alpha^{2} u \overline{u}), i = 1,2
\end{equation}

The axial symmetry of the string fields implies $\partial_{1} u \partial_{1} \overline{u}= \partial_{2}u \partial_{2}\overline{u}$.  With the use of this relation, we have from Eqs. (6) and (7)

\begin{equation}\label{Eq11}
T_{11}= T_{22}= \frac{\eta^{2}}{(1 + |u|^{2})^{2}}(\partial_{t}u \partial_{t}\overline{u} - \alpha^{2}u \overline{u})
\end{equation}

Using Eqs. (9-11) we obtain

\begin{equation}\label{Eq12}
T_{00}- \frac{1}{2}T = \frac{\eta^{2}}{(1 + |u|^{2})^{2}} (2 \partial_{t}u \partial_{t}\overline{u} - \alpha^{2}u \overline{u})
\end{equation}

Introducing this result and Eq. (8) in the 00-component of Eq. (3), we obtain 

\begin{eqnarray}\label{Eq13}
\nabla^{2} h_{00} = \frac{16 \pi G \eta^{2}\alpha^{2}u \overline{u}}{(1 + |u| ^{2})^{2}}
\end{eqnarray}

We have used the identity $\partial_{t}u \partial_{t}\overline{u} = \alpha^{2} u\overline{u}$ which follows from the solution$^{4}$

\begin{equation}\label{Eq14}
u(t,z) = \exp ( -i \alpha t) u (z)
\end{equation}

where $u (z)$ is a degree $N$ rational map in $z=x^{1}+ ix^{2}$

\begin{equation}\label{Eq15}
u (z) = \Bigr(\frac{\lambda}{z}\Bigr) ^{N} 
\end{equation}

Leese$^{4}$ finds that in order to have a lump of finite energy, the integer $N$ must be equal or larger than 2.  Incidentally, the same condition applies to the nonrelativistic spinning soliton.$^{9}$  This can be understood by considering the energy per unit string length 

\begin{equation}\label{Eq16}
\epsilon = \int d^{2}x T_{00} = 2 \pi \eta^{2} N + \frac{2 \pi^{2} \eta^{2}\alpha^{2}\lambda^{2}}{N^{2} \sin \frac{\pi}{N}}  
\end{equation}

where the second equality follows by integrating Eq. (9) with $u(t,z)$ given by Eqs. (14) and (15). 

We see that the second term in Eq. (16) diverges for $N=1$.  Since this term originates from the anisotropy energy, it is of no surprise to have the same condition for the nonrelativistic spinning soliton$^{9}$.

For the Q-lump, this term can be expressed by means of the Noether charge$^{4}$ $Q$

\begin{equation}\label{Eq17}
Q=i \int d^{2}x \frac{\overline{u}\partial_{t}u - u \partial_{t}\overline{u}}{(1+|u|^{2})^{2}}= 2\alpha \int d^{2}x \frac{u \overline{u}}{(1+|u|^{2})^{2}}
\end{equation}

With use of this result, the energy given in Eq. (16) becomes$^{10}$  

\begin{equation}\label{Eq.18}
\epsilon = \eta^{2} (2 \pi N + |\alpha Q|) 
\end{equation}

Starting from Eq. (13), we now proceed to the evaluation of the gravitational acceleration $\vec{g}(r)$.  Noting that $h_{00} = 2 \phi_{g}$ where $\phi _{g}$ is the gravitational potential related to the acceleration by $\vec{g} (r)= - \vec{\nabla} \phi_{g}$, Eq. (13) can be cast into the form of a Poisson equation

\begin{equation}\label{Eq.19}
\vec{\nabla}. \vec{g}(r) = - 4 \pi G \rho (r)
\end{equation}

where $\rho(r)$ is the active gravitational mass density (per unit volume) 

\begin{equation}\label{Eq.20}
\rho (r) = \frac{2 \eta^{2}\alpha^{2}\lambda^{2N} r ^{2N}}{(r^{2N} + \lambda^{2N})^{2} }
\end{equation}

Using Gauss theorem, Eq. (19) yields $g (r)$ (not to be confused with the determinant of the metric)

\begin{equation}\label{Eq21}
g (r) = - \frac{4 \pi G}{r} \int^{r}_{o} dr ' r ' \rho (r ')
\end{equation}

For $N=2$, we obtain from Eqs. (20) and (21)

\begin{eqnarray}\label{Eq22}
g (r)= 2 \pi G \eta^{2} \alpha^{2}\lambda^{4} (\frac{r}{r^{4}+ \lambda^{4}} - \frac{1}{\lambda^{2}r} \tan ^{-1} \frac{r^{2}}{\lambda^{2}})
\end{eqnarray}

This function has a maximum at $r \simeq \frac{1}{2}\lambda$.  Since $\lambda$ is the width of the sigma-model string, it is useful to have asymptotic results deep in the interior region, $r <<\lambda$, and in the exterior region, $r >> \lambda$. From Eq. (22), we obtain

\begin{equation}\label{Eq23}
g (r) \rightarrow -  \frac{4 \pi G \eta^{2}\alpha^{2}r^{5}}{3 \lambda^{4}}  
\end{equation} 

for $r/\lambda \rightarrow 0$ and   

\begin{equation}\label{Eq24}
g (r)\rightarrow - \frac{\pi^{2} G \eta^{2} \alpha^{2} \lambda^{2}}{r}
\end{equation}

for $r/\lambda \rightarrow \infty$. From Eq. (23), we see that $g(r)$ is a non-singular function of $r$ as we approach the string axis.  Introducing the linear active mass density $m$ (obtained for $N=2$ from Eq. (20))

\begin{equation}\label{Eq25}
m = 2\pi \int^{\infty}_{0} dr r \rho (r) = \frac {\pi^{2}\eta^{2} \alpha^{2} \lambda^{2}} {2} 
\end{equation}

we can express Eq. (24) as follows

\begin{equation}\label{Eq26}
g (r) \rightarrow - \frac{2 G m}{r} = - \frac{G \eta^{2} |\alpha Q|}{r}
\end{equation}

This result shows that, for $r$ well beyond the soliton width $\lambda$, the string acts as a massive rod of linear active mass density $m$.  Recalling Eq. (18), we see that $m = \eta^{2}|\alpha Q|$.   

\section{Induced Frame Dragging}

The dragging of inertial frames relative to the asymptotic frame inside a rotating mass shell has been first investigated by Thirring and Lense$^{6}$.  We expect that this effect is also present for a spinning string.  Calculation of the mixed space-time components of the metric tensor (2), described below confirms this expectation.

First consider the component $g_{01}(\vec{x})$.  Taking the $01$-component of Eq. (3) and using the weak field result $R _{01} = - \frac{1}{2} \nabla^{2} g _{01}$, we have

\begin{equation}\label{Eq27}
\nabla^{2}g_{01}(\vec{x}) = 16 \pi G T_{01} (\vec{x})
\end{equation}

From Eq. (7), we obtain in the weak field approximation

\begin{equation}\label{Eq28}
T_{01} = \frac{\eta^{2}}{(1 + |u|^{2})^{2}}(\partial_{0} u \partial_{1} \overline{u} + \partial_{0} \overline{u} \partial_{1} u)
\end{equation}

Substituting for $u(t, z)$ the solution (14), Eq. (28) yields for $N = 2$

\begin{equation}\label{Eq29}
T_{01}(\vec{x}) = - \frac{4 \eta^{2} \alpha \lambda^{4} r^{3}}{(r^{4} + \lambda^{4})^{2}}\sin\theta
\end{equation}

where we introduced polar coordinates $x^{1} = r \cos \theta$ and $x^{2} = r \sin \theta$.  To solve Eq. (27), we use the two dimensional Green's function

\begin{eqnarray}\label{Eq30}
G (\vec{x}- \vec{x}') = \frac{1}{2 \pi} \Bigg
{\{} \log r - \sum^{\infty}_{n=1}\ \frac{1}{n} \bigl(\frac{r '}{r}\bigr)^{n} \cos [n (\theta - \theta ')]\Bigg{\}} S (r-r') \nonumber \\* + \frac{1}{2 \pi} \Bigg{\{}\log r ' -\sum^{\infty}_{n = 1} \frac{1}{n} \bigl(\frac{r}{r'}\bigr)^{n} \cos [n (\theta - \theta ')]\Bigg{\}} S (r' - r)
\end{eqnarray}

where $S(r)$ is the unit step function.  Using Eqs. (29) and (30), the solution of Eq. (27) is written as follows

\begin{eqnarray}\label{Eq31}
g_{01} {\vec{(x)}} = \frac{\Gamma}{2 \pi}\int^{\infty}_{0} dr' r' F (r') \int^{2 \pi}_{0} d  \theta^{'} \sin \theta^{'}\Bigl[\frac{r'}{r} \cos (\theta - \theta^{'}) S (r - r') \nonumber \\*+ \frac{r}{r'} \cos (\theta - \theta^{'}) S (r' - r) \Bigr]
\end{eqnarray}

where $\Gamma = 64 \pi G \eta^{2}\alpha \lambda^{4}$, and $F (r) = r ^{3}(r^{4} + \lambda^{4})^{-2}$.  Performing the angular integrations, Eq. (31) yields

\begin{eqnarray}\label{Eq32}
g_{01}(\vec{x}) = \frac{1}{2} \Gamma \sin \theta
\Bigl[\frac{1}{r} \int^{r}_{0} dr' F (r')r '^{2} + r \int^{\infty}_{r} dr'F (r')\Bigr] \nonumber \\* = 8 \pi G \eta^{2} \alpha \lambda^{2} \frac{\sin \theta}{r}\tan^{-1} \frac{r^{2}}{\lambda^{2}}  
\end{eqnarray}

In a similar way, we calculate $g_{02} (\vec{x})$.  Solving the 02-component of Eq. (3), we have

\begin{equation}\label{Eq33}
g_{02} (\vec{x}) = - 8 \pi G \eta ^{2} \alpha \lambda^{2}\frac{\cos \theta}{r} \tan^{-1} \frac{r^{2}}{\lambda^{2}}  
\end{equation}

Using Eqs. (32) and (33), we calculate the local angular velocity, $\omega(r)$, of the frame rotation induced by the spinning of the Q-lump string.  This is done by expressing the third term of Eq. (2) in polar coordinates.  Using Eqs. (32) and (33), we have

\begin{equation}\label{Eq34}
2 [g_{01}(\vec{x})dx^{1} + g_{02}(\vec{x}) dx^{2}] dx^{0} = -16 \pi G \eta^{2} \alpha \lambda^{2} d \theta d x ^{0}\tan^{-1} \frac{r^{2}}{\lambda^{2}}
\end{equation}

On comparing the rhs of this equation with the cross term of the rotating metric component $r^{2}[d \theta - \omega (r) dx^{0}]^{2}$, we obtain

\begin{equation}\label{Eq35}
\omega (r) = 8 \pi G \eta^{2}\alpha (\frac{\lambda}{r})^{2} \tan^{-1} \frac{r^{2}}{\lambda^{2}} 
\end{equation}

Interesting conclusions emerge from the asymptotic expansions of (35). For $r/\lambda \rightarrow 0$, we have

\begin{equation}\label{Eq36}
\omega (r) \simeq 8 \pi G \eta^{2} \alpha = 2 \epsilon_{0} G \omega_{s}
\end{equation}

where $\epsilon_{0}= 4 \pi \eta^{2}$ is the linear mass density of the pure sigma model$^{5}$ given by the first term of Eq. (18).  We have also set $\alpha = \omega_{s}$.  Now, the rhs of Eq. (36) allows us to make a comparison with the well known Thirring$^{7}$ result for the rotation rate $\Omega (r)$ of the inertial frame induced by a thin spherical shell of radius $r_{0}$ and mass $M$, rotating with the angular velocity $\omega_{s}$.  In the interior region, $r<r_{0}$, $\Omega (r) = \Omega_{int} (r)$ is given by $^{7}$

\begin{equation}\label{Eq37}
\Omega_{int} (r) = \frac{4M G \omega_{s}}{3 r_{0}}
\end{equation}

Eq. (37) exhibits a similarity with the rhs of Eq. (36), except for the geometric factor $2/3 r_{0}$ characteristic of the spherical symmetry.  For $r/\lambda \rightarrow \infty$, we have from Eq. (35)

\begin{equation}\label{Eq38}
\omega (r) \simeq \pi \epsilon_{0} G \omega_{s} \frac{\lambda^{2}}{r^{2}}
\end{equation}

This is to be compared with the Thirring formula$^{7}$ for the exterior region, $r > r_{0}$, 

\begin{equation}\label{Eq39}
\Omega_{ext}(r) = \frac{ 4M G \omega_{s}}{3} \frac{r^{2}_{0}}{r^{3}}
\end{equation}

This time, the geometric factor $4/3 \pi r$ enters in Eq. (39).  Moreover, the comparison suggests that $\lambda$ in Eq. (38) plays a role of the radius of the rotating mass shell.  This view is substantiated by looking at the $r$ dependence of the mass density $T_{00}(r)$ of the pure sigma model.  According to Eq. (9), we have for $N=2$

\begin{equation}\label{Eq40}
T_{00} (r) = \frac{\eta^{2} \partial_{i} u \partial_{i} \overline{u}}{(1 + |u|^{2} )^{2}} = \frac {8 \eta^{2} \lambda^{4} r ^{2}}{(r^{4} + \lambda^{4})^{2}}
\end{equation}

From this expression we see that $T_{00}(r)$ goes as $r^{2}$, and $r^{-6}$ for $r/ \lambda \rightarrow 0$, and $r/\lambda \rightarrow \infty$, respectively.  The maximum of $T_{00}(r)$ is at $r \simeq 0.75 \lambda$.  In view of this comparison, we conjecture that the inertial frame dragging effect for the Q-lump string is similar to that for the cylindrical shell of radius of order $\lambda$, rotating at the angular velocity $\omega_{s} = \alpha$.

This conjecture can be substantiated quantitatively by computing the x$^{3}$ component, $J_{3}$, of the angular momentum per unit length of the string.  Using the expression $ \epsilon _{jk} x _{j} T_{0k}$ for the angular momentum density, we obtain in the weak field approximation $J_{3} \simeq \pi ^{2}\alpha \eta^{2} \lambda^{2}$.  This should be compared with the angular momentum, $J = 4 \pi \alpha \eta^{2} r _{0}^{2}$, of a thin cylindrical shell of radius $r_{0}$ with linear mass density, $\epsilon_{0}$, rotating at angular velocity $\alpha$.  From this comparison, we have $r_{0}^{2} \simeq \pi \lambda^{2}/4$ confirming the above conjecture. Note  that the rhs of Eq. (38) can be expressed in terms of $J$ yielding $\omega (r)\simeq 4 J G / r ^{2}$.  This is a string analog of the Thirring formula (39) expressed via the angular momentum of the spherical shell, $J _{sph} = 2 \omega _{s} M r_{0}^{2}/ 3$, as  $\Omega _{ext} (r) = 2 J_{sph} G/r^{3}$.

\section{Deflection of Light}

We assume that the light path is in the $(x^{1}, x ^{2})$ plane.  First we neglect the contribution to the deflection caused by the frame dragging.  Then the relevant metric is

\begin{equation}\label{41}
ds^{2} = g_{00} (dx^{0})^{2} + g_{11} [(dx^{1})^{2} + (dx^{2})^{2}]
\end{equation}

where

\begin{equation}\label{42}
g_{00} = 1 + 2 \phi , g_{11} = -1 + 2 \overline {\phi}
\end{equation}

The gravitational potentials $\phi = \frac{1}{2}h_{00}$ and $\overline{\phi} = \frac{1}{2} h_{11}$ are obtained by solving the 00 and 11-component of the Einstein equation (3), respectively.

To calculate the deflection angle, we consider the geodesic equation of motion for the four momentum, $p_{i}$, of the photon$^{12}$

\begin{equation}\label{43}
\frac{dp_{i}}{dt} = \frac{1}{2} g_{jk,i} \frac{p^{j}p^{k}}{p^{0}}
\end{equation}

Assuming that the light ray is approaching the string in the $y$ direction, the deflection is produced by the rate of increase of the $x$ component of the coordinate velocity $u^{x} = p^{x}/p^{0}$.  Thus, we consider Eq. (43) for $i = x$, and obtain with use of Eqs. (41) and (42)

\begin{equation}\label{44}
- \frac{d}{dt}\Big{[}(1 - 2 \overline{\phi})p^{x}\Big{]}= p^{0}\Big{[}\partial_{x}\phi + \partial_{x} \overline {\phi} \frac{(p^{x})^{2} + (p^{y})^{2}}{(p^{0})^{2}}\Big{]}
\end{equation}

Taking $i=0$, and noting that the metric tensor $g_{jk}$ is independent of time, equation (43) yields

\begin{equation}\label{45}
\frac{d}{dt} [(1 + 2 \phi) p^{0}] = 0
\end{equation}

Using this result, it follows that to first order in $\phi$ and $\overline{\phi}$, the following identity holds

\begin{equation}\label{46}
\frac{1}{p^{0}} \frac{d}{dt} \Big{[}(1 - 2 \overline {\phi}) p^{x}\Big{]} \simeq \frac{d}{dt}\Big{[}(1 - 2 \overline{\phi} - 2 \phi) \frac{p^{x}}{p^{0}}\Big{]}
\end{equation}

Further simplification of Eq. (44) follows by noting that $p^{i}$ is a null vector.  With the metric (41), this implies

\begin{equation}\label{47}
\frac{(p^{x})^{2} + ( p^{y})^{2}}{(p^{0})2} = \frac{1 + 2 \phi}{1 - 2 \overline {\phi}} \simeq 1
\end{equation}

Dividing Eq. (44) by $p^{0}$, we obtain with use of Eqs. (46) and (47) to order $\phi$, $\overline \phi$

\begin{equation}\label{48}
\frac{d u ^{x}}{dt} \simeq - \partial_{x} [ \phi (r) + \overline {\phi} (r)] = - \frac{x}{r} \partial _{r} [\phi(r) + \overline {\phi} (r)]
\end{equation}

From Eq. (22), we have $\partial_{r} \phi = \partial_{r} \phi _{g} = - g (r)$.  To obtain the quantity $\partial_{r} \overline {\phi}$, we consider the Einstein equation (3) for $h_{11}$

\begin{equation}\label{49}
\nabla^{2} h_{11} = 16 \pi G (T_{11} + \frac{1}{2} T) = \frac{16 \pi G \eta^{2}\lambda^{4}}{(r^{4} + \lambda^{4})^{2}} (8 r ^{2} + \alpha^{2} r^{4})
\end{equation}

where the second equality follows from Eqs. (9) and (11) using the ansatz (15) with $N=2$.  Similar to the derivation of Eq. (22), we use Gauss theorem to solve Eq. (49) for $\partial_{r} h_{11}$.  Combining this result with Eq.(22), we have

\begin{equation}\label{50}
\partial_{r} (\phi + \overline{\phi}) = 4 \pi G \eta^{2} \Bigl{[}\frac{4 r^{3}}{r^{4} + \lambda^{4}} + \alpha^{2}\lambda^{2}\Bigl{(} \frac{1}{r}\tan ^{-1}\frac{r^{2}}{\lambda^{2}} - \frac{\lambda^{2}r}{r^{4} + \lambda^{4}}\Bigr{)} \Bigr{]}
\end{equation}

The differential of the deflection angle $\delta\Phi$, acquired during time interval $dt$ is $d (\delta \Phi) = (du^{x}/dt)dy$.  Using Eq. (48), the net deflection angle $\delta \Phi$ becomes 

\begin{equation}\label{51}
\delta \Phi = - x \int^{\infty}_{- \infty} dy \frac{1}{r} \partial_{r} (\phi + \overline\phi)
\end{equation}

where $r = [x^{2} + y ^{2}]^{\frac{1}{2}}$ and $x$ is the impact parameter.  In what follows, we assume $x \gg \lambda$ implying that also $r \gg \lambda$.  In this limit, Eq. (50) yields

\begin{equation}\label{52}
\partial_{r} (\phi + \overline {\phi}) \approx \frac{\epsilon_{0}G} {r} (4 + \frac{\pi}{2} \alpha^{2} \lambda^{2})
\end{equation}

Using this result in Eq. (51), the deflection angle becomes

\begin{equation}\label{53}
\delta \Phi \approx  - x \epsilon_{0} G (4 + \frac{\pi}{2} \alpha^{2} \lambda^{2}) \int^{\infty}_{-\infty} \frac{dy}{(x^{2} + y^{2})} = - \pi \epsilon_{0} G (4 + \frac{\pi}{2} \alpha^{2} \lambda^{2})
\end{equation}

We see that in the limit $x/\lambda \rightarrow \infty$, the deflection angle is independent of $x$.  Moreover, the rhs of Eq. (53) can be written as $\delta \Phi \approx - 4 \pi G \epsilon$, where $\epsilon $ is the net energy per unit length given in Eq. (16).  These results are reminiscent of the deflection angle obtained from the wedge angle deficit of the conical space$^{3,5}$.  In fact, for $\alpha = 0$, the angle $2 \delta \Phi \approx -8 \pi G \epsilon_{0}$ coincides with the deficit angle $\delta$ obtained in Ref. 5 using the Gauss-Bonet Bonnet formula.

We now consider the additional contribution to light deflection due to the frame dragging.  Denoting the corresponding deflection angle as $\delta \Phi_{d}$, we obtain with use of Eq. (35)

\begin{equation}\label{54}
\delta \Phi_{d} = \int^{\infty}_{-\infty} dy \omega (r) = 8 \pi G \eta^{2} \alpha \lambda^{2} \int ^{\infty}_{-\infty} dy \frac{\tan ^{-1}(\frac{r^{2}}{\lambda^{2}})}{r^{2}}
\end{equation}

An approximate evaluation of $\delta \Phi_{d}(x)$ for $0 < x < \infty$ can be made by replacing the integrand of Eq. (54) by $(\pi/2) (r^{2} + \pi \lambda^{2}/2)^{-1}$. In this way, we get

\begin{equation}\label{55}
\delta \Phi_{d}(x) \sim \frac{\pi^{2}\epsilon_{0}G \alpha \lambda^{2}}{\sqrt{x^{2}+ \frac{\pi \lambda^{2}}{2}}}
\end{equation}

\section{Discussion}

According to Eq. (25), an infinite stationary Q-lump spinning string acquires an active linear mass density $m = \frac{\pi}{8} \epsilon_{0} (\alpha \lambda)^{2}$ where $\epsilon_{0}$ is the linear mass density of pure sigma model string.  The fact that matter can be attracted onto these strings prompts us to examine their role in the formation of galaxies.

The main concern is the magnitude of the density inhomogeneity due to infinite strings, $\delta \rho / \rho = \rho_{inf} / \rho$.  According to Ref. 2, $\rho _{inf}$ at cosmic time $t$ is given by $m/t^{2}$.  In the radiation dominated era we have $\rho = 3/32 \pi G t^{2}$.  Thus

\begin{equation}\label{56}
\frac{\rho _{inf}}{\rho} = \frac{32 \pi}{3} G m
\end{equation}

Galaxy formation scenarios require that $G m \sim 10^{-6}$.  Let us first consider the pure sigma model string with linear mass density $\epsilon_{0} = 4 \pi \eta^{2}$.  These strings are formed when the universe cools to temperature $T \sim \eta$.  We note that at this temperature the Higgs field acquires a nonzero vacuum expectation value owing to the O (3) invariant Higgs potential.  With $m$ replaced by $\epsilon_{0}$, the condition $G \epsilon_{0} \sim 10 ^{-6}$ implies $\eta \sim 3 \times 10^{15}$ Gev which falls into the grand unification (GU) region.

Although static sigma model string does not accrete matter, it can produce gravitational acceleration, when in motion.  Formation of planar wakes behind a relativistically moving string is one possible mechanism for the generation of the density fluctuations.  Another alternative is matter accretion onto oscillating closed loops$^{2}$.

For the stationary Q-lump spinning string we have $G m = G\epsilon_{0} \pi (\alpha \lambda)^{2}/{8}$.  Since the formation of galaxies requires $Gm \sim 10^{-6}$, we get a condition $\alpha \lambda \sim 1$ if $G \epsilon_{0} \sim 10^{-6}$ is assumed.

The Q-lump string can be detected through gravitational deflection of light.  When the impact parameter is large so that $x >> \lambda$, Eq. (53) shows that the deflection angle is independent of $x$ and its magnitude $\delta \Phi \simeq 4 \pi G \epsilon$ where $\epsilon$ is the net energy per unit length of the string.  In the limit $ \alpha \lambda \rightarrow 0$, these features agree with the deflection angle for vacuum string obtained in Ref. 3.  For $\alpha \lambda \sim 1$, there is a considerable enhancement (about 40$\%$) of the deflection angle due to spinning. 

Eq. (55) shows that the drag-induced light deflection angle depends not only on the product $\alpha \lambda$ but also on the string radius $\lambda$ itself.  However, in the Q-lump model, the radius is not fixed (being determined by the Noether charge Q which is a free parameter).  An orientational estimate of $\lambda$ can be made with use of a related string model that is based on a spinning Baby Skyrmion model of Piette et al$^{13}$.  In this model, the soliton has a preferred size determined by competition between the potential and Skyrme terms in the Lagrangian.  Recently, we studied the gravitational field of a string obtained by trivially extending the Baby Skymion model from $2 + 1$ to $3+1$ dimensions$^{14}$.  We note that the spinning frequency, $\omega$, of this model is a free parameter.  However, by requiring that the active gravitational mass be a relevant factor in galaxy formation, the product $\omega \lambda$ must be of order of $1$.  Also, to ensure exponential localization of the soliton, $\omega$  must be smaller than the magnitude of the potential term.  From these conditions, we deduce that $\lambda < 1/ \eta$.  Thus, the radius of a GU string is comparable to Higgs Compton wavelength $\lambda \sim 10^{-31}$ meter.  We note that this result is similar to the transverse dimensions of strings studied in Ref. 3. With this value of $\lambda$, we now proceed to make an estimate of $\delta \Phi_{d}$ from Eq. (55).  Denoting by $R$ the proper distance of the observer to the string axis, the impact parameter $x \sim R \delta \Phi_{d}$.  Owing to the extremely small value of $\lambda$, the second term in the denominator of Eq. (55) can be neglected yielding $\delta \Phi_{d}\sim (\pi^{2}\epsilon_{0} G \alpha \lambda^{2}/R)^{\frac{1}{2}}$.  Taking $\epsilon_{0} G \sim 10 ^{-6}, \alpha \lambda \sim 1, \lambda \sim 10 ^{-31} $meter, and $R \sim 10^{4} $ meter, we get $\delta \Phi_{d} \sim 10 ^{-20} $ rad which is negligible compared with the deflection angle $\delta \Phi \sim 10 ^{-5}$ resulting from Eq. (53).

Now let us turn to some problems of the dynamics of spinning strings.  First note that a string can remain static only if it is straight.  Curved strings oscillate under their own tension.  Of particular importance for the galaxy formation are oscillating closed loops$^{2}$.  In this case, the oscillation of the string axis is responsible for the generation of an active gravitational mass that bears qualitative similarities with the mechanism outlined in Sec. 3.  This may be seen by recalling Eq. (25), which shows that the active mass induced in a unit length of a spinning string is of the order of $\eta^{2}$ times the square of the velocity at the radius $\lambda$ of the cylindrical shell.  This should be compared with the mass $M \sim \eta^{2} v ^{2}_{rms}$ derived by Turok$^{15}$ for a loop oscillating with rms velocity $v_{rms}$.

Now, the question that needs to be addressed is if closed loops of spinning strings could also serve as seeds for galaxy formation. The scenario of galaxy formation proposed by Vilenkin$^{2}$ requires, that the main energy loss mechanism of large loops is graviational radiation.  For gauge-symmetry non-spinning strings, Vachaspati et al$^{16}$ have shown that electromagnetic radiation and the radiation of massive particles fails to yield significant energy loss in comparison to the gravitaional radiation.  For the Q-lump string, considered in this paper, there is no gauge field in the Lagrangian density of Eq. (1).  Thus, there is no coupling of the photon field to the oscillations of the string as a whole.  However, there is non-zero coupling of these oscillations to the field corresponding to small oscillations, $\delta \vec{\phi}$, of the vector $\vec{\phi}$ about the vacuum configuration $\vec{\phi}_{0} = (0, 0, 1)$.  This coupling is of similar nature as that considered by Davis$^{17}$for global non-spinning string.  Specifically, the $\delta \vec{\phi}$ field is rigidly carried along with the string as it moves.  Nevertheless,  the power radiated due to this coupling is zero as long as the frequency of the string oscillation $\omega$ is less than the spinning frequency $\alpha$.  This is because $\delta \vec{\phi}$ satisfies a Klein-Gordon equation describing massive bosons with the mass $\alpha$.  For a loop of size $R, \omega \sim 1/R$, whereas $\alpha$ is of order $\eta \sim 10^{15}$ GeV.  Thus, $\alpha \gg \omega$, implying that the decay into the $\delta \vec {\phi}$ bosons is ruled out.  This contrasts with the case of global strings resulting from a spontaneously broken U(1) gauge theory $^{17}$.  In this case, the decay is into massless bosons corresponding to oscillations of the phase of the scalar field resulting in a lifetime of large oscillating loop that is much shorter than that due to the gravitational radiation.

Thus we come to the conclusion that owing to the presence of the anisotropy-induced boson mass the gravitational radiation is the main mechanism of energy loss for Q-lump string.  In this context, it would be interesting to study gauged $\sigma$-model strings with Chern-Simons term$^{18}$.  Such strings are expected to be electrically charged and their oscillation could possibly couple to the field of a single photon.  Then electro-magnetic radiation may become a significant factor for energy loss.

\section{Summary}

We have studied the gravitational field of an infinite straight and stationary spinning string.  By including into the pure sigma model string$^{5}$ an uniaxial anisotropic energy, we arrive at the model of a Q-lump string.  The 2+1 dimensional version of this model was first studied by Leese$^{4}$.  The time dependent solutions of this model describe spinning in the internal space.  The kinetic energy associated with this spinning disturbs the perfect balance of the string tension and the positive energy originally responsible for the absence of long range gravitational field of non-spinning strings$^{3,5}$.  The approximate solution of the Einstein equations shows that a spinning string exhibits long range acceleration resembling that of a rotating massive cylindrical shell.  We also study the dragging of the inertial frames caused by the spinning.  The results for the angular velocity of the induced dragging confirm the picture according to which a spinning string of degree $N=2$ behaves like a rotating cylindrical mass shell of radius given by the string width. 

Finally, we study the deflection of a light beam propagating in the plane perpendicular to the axis of spinning string.  If the contribution of the frame dragging is neglected, the deflection angle is given by $4 \pi G \epsilon$ where $\epsilon$ is the net energy per unit length of the string.  This result is consistent with the deficit angle of the conical space obtained for the pure sigma model in Ref. [5] using the Gauss-Bonnet formula.  Owing to the fact, that the string width is extremely small (being given by the Compton wavelength of the GU Higgs boson), the contribution of the frame dragging to the deflection angle is negligible.

The central result of this work is Eq. (25) for the linear active mass density.  Using this formula, we conclude that straight Q-lump strings could serve as seeds for galaxy formation if the product of the angular velocity times the string width is comparable to the speed of light.  Moreover, closed oscillating loops of spinning strings satisfy the condition that their energy loss is dominated by gravitational radiation as required by Vilenkin scenario$^{2}$.

{\em Note added.} When I wrote this paper I was not aware of several of related earlier papers.  Thanks to correspondence from Professors S. Deser, R. Jackiw, and Y. Verbin I have been enlightened and some of the most relevant papers are now included as Refs. [19]-[24].

The global properties of the 2 + 1-dimensional space-times generated by massive point particle with angular momentum $J$ have been thoroughly investigated in Ref. [19].  It is of interest to note the relation of the line element given in Eq. (4. 17) of this reference to Eq. (34) of the present paper.  As $r/\lambda \rightarrow \infty$, the rhs of this equation goes to $-8G J_{3} d \theta dx^{0}$, where $J_{3}$ is the $x^{3}$-component of the angular momentum per unit length of the Q-lump string.  This result should be compared with the term $2A dt d \theta = - 8 GJ dt d \theta$ in Eq. (4.17).  This agreement is not surprising, since we show that the spinning string behaves as a rotating cylindrical mass shell (see Sec 4).  In ref. [20] and [21], the 2+1 dimensional metric of Ref. [19] has been extended to study the gravitational effects of straight spinning string.  The physical consequences  of the "time-helical" structure of the locally flat coordinates derived in Ref. [19] have been thoroughly studied in Ref. [23] which presents solutions of the Klein-Gordon and Dirac equations in the presence of massive point particle with arbitrary angular momentum.

Closer to our present work appears to be a more recent paper by Verbin and Larsen [24] who study the Q-lump string with a general oscillatory behavior of the fields given by $\exp (iqz-\omega t)$.  The case $q=0$ corresponds to the present model, but since Ref. [24] goes beyond the weak coupling approximation, the solutions for the metric are obtained numerically.  From Fig 6. of Ref. [24] we see that the metric $N = 1 + h_{00}$ is consistent with our result showing that $h_{00}$ goes as $(r/\lambda)^{6}$ for $r/\lambda \rightarrow 0$ and as $\log (r/\lambda)$ for $r/\lambda \rightarrow \infty$.  Thus the metric $g_{00}$ is not asymptotically flat owing to finite frequency of spinning.  This general trend is also exhibited by other examples studied numerically by Verbin and Larsen [24].  This confirms predictions made by these authors following a deep analysis based on Kaluza-Klein reduction from $D$-dimensional global strings to $D-1$-dimensional gauged strings.  On the other hand, the numerically obtained quantity $L_{\varphi} = 4GJ_{3}$ is asymptotically flat in agreement with Eqs. (32) and (33) of our work.

% Set the ending of a LaTeX document

\begin{thebibliography}{17}

\bibitem{1}
T.~W.~B. Kibble,
\emph{}
J. Phys. A \textbf{9}, 1387 (1976).

\bibitem{2}
A. Vilenkin,
\emph{}
Phys. Rep. \textbf{121}, 263 (1985).

\bibitem{3}
A. Vilenkin, 
\emph{}
Phys. Rev. D \textbf{23}, 852 (1981).

\bibitem{4}
R.~A. Leese, 
\emph{}
Nucl. Phys. B \textbf{366}, 283 (1991).

\bibitem{5}
A. Comtet and G.~W. Gibbons, 
\emph{}
Nucl. Phys. B \textbf{299}, 719 (1988).

\bibitem{6}
H. Thirring and J. Lense, 
\emph{}
Phys. Z. \textbf{19}, 156 (1918).

\bibitem{7}
H. Thirring,
\emph{}
Phys. Z. \textbf{19}, 33 (1918); \textbf{22}, 29 (1921).

\bibitem{8}
D.~R. Brill and J.~M. Cohen, 
\emph{}
Phys. Rev. \textbf{143}, 1011 (1966).

\bibitem{9}
A.~M. Kosevich,
in \emph {Solitons}, edited by S. E. Trullinger, V. E. Zakharov, and V. L. Pokrovsky
(Elsevier Science Publishers, B. V., 1986).

\bibitem{10}
N. Manton and P. Sutcliffe,
\emph {Topological Solitons},
(Cambridge Monographs on Mathematical Physics, Cambridge, 2004).

\bibitem{11}
A. Einstein,
\emph {The Meaning of Relativity},
(Princeton University Press, Princeton, Newy Jersey, 1972).

\bibitem{12}
P.~J.~E. Peebles, 
\emph {Principles of Physical Cosmology},
(Princeton University Press, Princeton, New Jersey, 1993.

\bibitem{13}
B.~M.~A.~G. Piette, B.~J. Schroers, and W.~J. Zakrzewski, 
\emph{}
Nucl. Phys. B \textbf{439}, 205 (1995).

\bibitem{14}
E. Simanek,
\emph {}
\emph{unpublished}.


\bibitem{15}
N. Turok, 
\emph{}
Nucl. Phys. B \textbf{242}, 520 (1984).


\bibitem{16}
T. Vachaspati, A.~E. Everett, and A. Vilenkin, 
\emph{}
Phys. Rev. D \textbf{30}, 2046 (1984).


\bibitem{17}
R.~L. Davis,  
\emph{}
Phys. Rev. D \textbf{32}, 3172 (1985).


\bibitem{18}
H.~J. de Vega, and F.~A. Schaposnik,  
\emph{}
Phys. Rev. Lett.  \textbf{56}, 2564 (1986).

\bibitem{19}
S. Deser, R. Jackiw, and G.~T. Hooft,  
\emph{}
Ann. Phys.  \textbf{152}, 220 (1984).

\bibitem{20}
P.~O. Mazur,  
\emph{}
Phys. Rev. Lett.  \textbf{57}, 929 (1986).

\bibitem{21}
D. Harari, and A.~P. Polychronakos,  
\emph{}
Phys. Rev. D  \textbf{38}, 3320 (1988).

\bibitem{22}
S. Deser, and R. Jackiw,  
\emph{}
Ann. Phys. \textbf{192}, 352 (1989).

\bibitem{23}
P. de Sousa, and R. Jackiw,  
\emph{}
Commun. Math. Phys.  \textbf{124}, 229 (1989).

\bibitem{24}
Y. Verbin, and A.~L. Larsen,  
\emph{}
Phys. Rev. D  \textbf{70}, 085004 (2004).





















\end{thebibliography}
\end{document}